\documentclass[preprint,floatfix] {revtex4} 
 
\usepackage{graphicx}
\usepackage{subfigure}
\begin{document}

\title{Studies on bound-state spectra of Manning-Rosen potential}
\author{Amlan K. Roy}
\altaffiliation{Email: akroy@iiserkol.ac.in, akroy6k@gmail.com, Ph: +91-3473-279137, Fax: +91-33-25873020.}
\affiliation{Division of Chemical Sciences,   
Indian Institute of Science Education and Research (IISER)-Kolkata, 
Mohanpur Campus, P. O. BCKV Campus Main Office, Nadia, 741252, WB, India.}

\begin{abstract}
Accurate ro-vibrational energies, eigenfunctions, radial densities, expectation values are presented for the exponential-type Manning-Rosen (MR)
potential. Bound states accurate up to ten significant figure are obtained by employing a simple, reliable generalized pseudospectral method. 
\emph{All} 55 eigenstates with $n \leq 10$ are treated for arbitrary values of potential parameters, covering a wide range of interaction, 
through a \emph{non-uniform, optimal} spatial radial discretization. A detailed investigation has been made
on energy changes with respect to \emph{screening and other} potential parameters. A systematic estimation of \emph{critical} screening parameters 
are given for these eigenstates. Special emphasis has been given to \emph{higher} states and in the 
vicinity of \emph{critical screening} region. A thorough comparison with literature results is made wherever possible. This \emph{surpasses} the 
accuracy of \emph{all} other existing methods currently available. Several \emph{new} states are reported for the first time. In short, 
a simple, efficient scheme for accurate calculation of this and other molecular potentials is offered. 


\end{abstract}
\maketitle

\section{Introduction}
The exponential-type Manning-Rosen (MR) potential \cite{manning33}, given by, 
\begin{equation}
v(r)= \frac{1}{\kappa b^2} \left[ \frac{\alpha (\alpha-1) e^{-2r/b}}{(1-e^{-r/b})^2} -
      \frac{Ae^{-r/b}}{1-e^{-r/b}}  \right], \ \ \ \ \ \kappa=\frac{2\mu}{\hbar^2}, 
\end{equation}
is used as an important mathematical model for molecular vibrations and rotations \cite{leroy70,cai86}. Here, $\alpha$ and  
the strength parameter $A$ are two dimensionless parameters, while the screening parameter $b$, having a 
dimension of length, is related to the range of potential. This has found considerable applications in several 
bound-state and scattering problems in physics. It is obvious that 
the potential remains invariant under the transformation $\alpha \leftrightarrow 1-\alpha$. For $\alpha=0$ or 1, this 
equation reduces to the familiar short-range Hulth\'en potential \cite{hulthen42} having useful applications in nuclear, 
particle, solid-state and chemical physics. There is a relative minimum value,  
$v(r_0)=-\frac{A^2}{4\kappa b^2 \alpha (\alpha-1)},$ at $r_0=b \ \mathrm{ln}[1+\frac{2\alpha (\alpha-1)}{A}],$ for 
$\alpha > 1$ and $A >0$.

This potential has received significant attention in recent years. It is well-known that the Schr\"odinger equation for 
$s$ states ($\ell = 0$)  of this potential can be solved \emph{exactly}. Such bound states are obtained analytically by a number of
attractive routes; \emph{viz.,} a direct factorization method \cite{infeld51}, Feynman path-integral formalism \cite{diaf05}, where 
eigenenergies, eigenfunctions are extracted from poles, residues of Green's function respectively, a standard
function analysis method expressing solutions in terms of generalized hyper-geometric functions \cite{dong07}, a 
tridiagonal matrix representation of wave operator in a complete square integrable basis \cite{mincang10}, etc.  
\emph{Exact} solutions of $s$-wave scattering states also are obtained from standard method \cite{chen07}.

However, the non-zero angular momentum states cannot be obtained \emph{exactly} analytically in closed form. Therefore, several 
approximation schemes have been proposed for these with varying degrees of accuracy and efficiency. The first definitive 
results for arbitrary $\ell$ states were presented by invoking an approximation of 
$\frac{1}{r^2} \approx \frac{1}{b^2} \frac{e^{-r/b}}{(1-e^{-r/b})^2}$ for the centrifugal term in short range 
\cite{qiang07}, quite similar in spirit to the familiar Pekeris approximation. Some of the other notable approaches are: 
a super-symmetric shape invariance formalism in conjunction with 
function analysis method \cite{chen09}, an approximation for centrifugal term different from the usual one used above, 
containing 3 adjustable parameters in it \cite{qiang09}, yet another alternative approximation to the centrifugal 
potential within a Nikiforov-Uvarov method \cite{ikhdair11}, the Duru-Kleinert method of path-integral formalism 
\cite{diaf11}, Laguerre and oscillator bases to tridiagonalize the
reference Hamiltonian and subsequently a Gauss quadrature approach for estimation of potential matrix elements \cite{hady11},
a J-matrix method \cite{nasser13}, etc. Approximate analytical scattering-state solutions of $\ell$-wave Schr\"odinger equation 
have been presented in \cite{wei08} by a proper approximation of the centrifugal term, as well as by the
J-matrix method \cite{nasser13}. Some other methods have also been presented \cite{gu11, arda12}. 
A purely numerical integrating procedure has also been programmed \cite{lucha99} as well for bound states, invoking the 
MATHEMATICA package, which offers decent results, especially in the short potential range, i.e., small $\ell$ and $\alpha$. 

Recently, the generalized pseudospectral (GPS) method has been shown to be quite successful for a number of physical situations, 
including the spiked harmonic oscillator, Hulth\'en, Yukawa, logarithmic, power-law, Hellmann, exponential-screened Coulomb
potentials, etc., as well as lower and higher states (including Rydberg states) in atoms and molecules (see the references
\cite{roy04,roy04b,roy05,roy05a,roy07,roy08,roy08a,roy11,roy13} and therein). Very accurate, reliable results were 
obtained through a \emph{non-uniform, optimal} spatial discretization in all these cases. In the present communication, our interest 
is to study the ro-vibrational spectra of MR potential through this GPS method, in order to assess its validity and performance in the
current context. That will help extend the domain of applicability of the method to a broader range of physical systems.
One interesting aspect of this potential is that there is a value of screening parameter, denoted by $(b^{-1})_c$, the 
\emph{critical screening parameter}, beyond which the state $(n,\ell)$ ceases to be a bound state. So far, only in one of the references 
\cite{nasser13}, some attention has been paid on this important issue; here we put particular emphasis to the eigenspectrum 
\emph{close to such threshold regions}. Moreover, excellent quality results are available for low-lying states, while only very few studies 
have been devoted to \emph{high-lying states}. We find interesting complex level crossing in higher $(n, \ell)$ states more predominantly 
than in lower states, in regions close to zero energy, which has remained hitherto unobserved. A thorough analysis on variation of energies with 
respect to screening parameter and $\alpha$ are presented. To this end, accurate ro-vibrational energies and wave functions of \emph{all} 
the 55 levels corresponding to $n \leq 10$ states of MR potential are reported. Screening parameters of arbitrary field strengths 
(covering both weak and strong limits of interaction) have been considered for particular values of vibrational and rotational 
quantum number. For further understanding, radial probability densities, 
expectation values are also reported for some selected states. A detailed comparison with literature results has been made 
wherever possible. The article is organized as follows. Section II gives a brief summary of the GPS method. A discussion of our
results is given in Section III, while Section IV makes a few concluding remarks. 

\section{GPS method for MR potential}
In this section, an overview of the employed methodology is presented. More details could be found in the references
\cite{roy04,roy04b,roy05,roy05a,roy07,roy08,roy08a,roy11,roy13} and therein. For the purpose of maintaining consistency with 
literature, we choose $A=2b$. Atomic units employed throughout the article, unless otherwise mentioned.

We are interested in the solution of radial Schr\"odinger equation, which can be written in following operator form, 
\begin{equation}
\hat{H}(r)\ \phi(r) =\varepsilon \ \psi(r).
\end{equation}
The Hamiltonian operator includes usual kinetic and potential energy terms,   
\begin{equation}
\hat{H}(r) =-\frac{1}{2} \ \ \frac{d^2}{dr^2} +v_{\mathrm{eff}}(r),
\end{equation}
with 
\begin{equation}
 v_{\mathrm{eff}}(r) = v(r) + \frac{\ell (\ell+1)}{2r^2} 
\end{equation}
and 
$v(r)$ is the MR potential, as given in Eq.~(1). The symbols have their usual significances. 

Majority of the finite-difference discretization schemes for solution of radial Schr\"odinger equation arising in these situations 
often require very large number of spatial grid points, mainly due to their uniform distributional nature. GPS method, 
on the other hand, facilitates a nonuniform, optimal discretization, maintaining similar kind of accuracies at both small as well as 
large $r$ regions. Therefore, one has the advantage of working with a much lesser grid points compared to many other methods in the 
literature. Thus we can have a denser mesh at smaller $r$ while a coarser mesh at large $r$. The principal feature of this scheme is 
that a function $f(x)$ defined in an interval $x \in [-1,1],$ can be approximated by a polynomial $f_N(x)$ of order N of the 
following form,   
\begin{equation}
f(x) \cong f_N(x) = \sum_{j=0}^{N} f(x_j)\ g_j(x).
\end{equation}
At the \emph {collocation points} $x_j$, above approximation is \emph{exact}, i.e., 
$ f_N(x_j) = f(x_j).$ In the Legendre pseudospectral method used in this study, $x_0=-1$, $x_N=1$, while
the $x_j (j=1,\ldots,N-1)$ are obtained from roots of first derivative of Legendre polynomial $P_N(x)$ with respect to  $x$, i.e., 
$ P'_N(x_j) = 0.$
The cardinal functions $g_j(x)$ in Eq.~(5) are given by the following expression,
\begin{equation}
g_j(x) = -\frac{1}{N(N+1)P_N(x_j)}\ \  \frac{(1-x^2)\ P'_N(x)}{x-x_j},
\end{equation}
satisfying a unique property that $g_j(x_{j'}) = \delta_{j'j}$. Now the semi-infinite domain $r \in [0, \infty]$ is mapped onto 
a finite domain $x \in [-1,1]$ by a transformation of the form $r=r(x)$. Next, one could use the following algebraic nonlinear 
mapping, $ r=r(x)=L\ \ \frac{1+x}{1-x+\eta},$ with L and $\eta=2L/r_{\mathrm{max}}$ being the two mapping parameters, as well a 
transformation of the form, $ \psi(r(x))=\sqrt{r'(x)} f(x)$. This, coupled with a symmetrization procedure, subsequently
leads to a transformed Hamiltonian as below, 
\begin{equation}
\hat{H}(x)= -\frac{1}{2} \ \frac{1}{r'(x)}\ \frac{d^2}{dx^2} \ \frac{1}{r'(x)}
+ v(r(x))+v_m(x),
\end{equation}
where $v_m(x)$ is given by the following relation,
\begin{equation}
v_m(x)=\frac {3(r'')^2-2r'''r'}{8(r')^4}.
\end{equation}
The advantage is that this leads to a \emph {symmetric} matrix eigenvalue problem which can be readily solved to produce accurate 
eigenvalues, eigenfunctions at the same time using standard available routines, quite easily and efficiently. This discretization 
then finally leads to a set of coupled equations as below, 
\begin{widetext}
\begin{equation}
\sum_{j=0}^N \left[ -\frac{1}{2} D^{(2)}_{j'j} + \delta_{j'j} \ v(r(x_j))
+\delta_{j'j}\ v_m(r(x_j))\right] A_j = EA_{j'},\ \ \ \ j=1,\ldots,N-1,
\end{equation}
\end{widetext}
where
\begin{equation}
A_j  = \left[ r'(x_j)\right]^{1/2} \psi(r(x_j))\ \left[ P_N(x_j)\right]^{-1},
\end{equation}
and $D^{(2)}_{j'j}$ is the symmetrized second derivative of cardinal function. For their expressions and other details, please
see the references \cite{roy04,roy04b,roy05,roy05a,roy07,roy08,roy08a,roy11,roy13} and therein.

In order to make a judicious choice of mapping parameters, a large number of tests were carried out to check the performance of this scheme. 
A sufficiently large range of potential parameters were scanned for this purpose to gain confidence. All our results are reported 
only up to the precision that were found to 
maintain stability with respect to these variations. In this way, a consistent set of numerical parameters ($r_{\mathrm{max}}=300,$ $\eta=25$ 
and $N=300$) has been chosen which seemed to be appropriate and satisfactory for the current problem. For higher excited states and 
also near the critical screening region, $r_{\mathrm{max}}$ was increased, for obvious reasons. These are mentioned in appropriate 
places in the text.

\begingroup
\squeezetable
\begin{table}
\caption {\label{tab:table1} Estimated critical screening parameters, $(b^{-1})_c$, of MR potential, for $\alpha=0.75$.} 
\begin{ruledtabular}
\begin{tabular}{l|llllllllll}
State  & \multicolumn{10}{c}{$n$}  \\
\cline{2-11} 
     &   1       &    2     &    3     &    4    &     5     &     6     &     7     &      8    &    9    &    10    \\
\hline 
 $s$ &  2.64192  & 0.61200  & 0.25709  & 0.13994 & 0.08769   & 0.06002   & 0.04363   & 0.03313   & 0.02601 & 0.02096  \\ 
 $p$ &           & 0.38428\footnotemark[1]  & 0.19043\footnotemark[1]  & 0.11264\footnotemark[1] & 0.07413\footnotemark[1]   & 0.05238\footnotemark[1]   
     & 0.03893\footnotemark[1]   & 0.03005\footnotemark[1]   & 0.02389 & 0.01944  \\  
 $d$ &           &          & 0.15866\footnotemark[1]  & 0.09828\footnotemark[1] & 0.06659\footnotemark[1]   & 0.04799\footnotemark[1]   
     & 0.03618\footnotemark[1]   & 0.02823\footnotemark[1]   & 0.02262 & 0.01853  \\ 
 $f$ &           &          &          & 0.08667\footnotemark[1] & 0.06019\footnotemark[1]   & 0.04414\footnotemark[1]   & 0.03370\footnotemark[1]   
     & 0.02655\footnotemark[1]   & 0.02144 & 0.01767  \\
 $g$ &           &          &          &         & 0.05460\footnotemark[1]   & 0.04067\footnotemark[1]   & 0.03142\footnotemark[1]   
     & 0.02498\footnotemark[1]   & 0.02032 & 0.01684  \\
 $h$ &           &          &          &         &           & 0.03754\footnotemark[1]   & 0.02932\footnotemark[1]   & 0.02351\footnotemark[1]   
     & 0.01925 & 0.01605  \\
 $i$ &           &          &          &         &           &           & 0.02740\footnotemark[1]   & 0.02214\footnotemark[1]   & 0.01825 & 0.01529  \\
 $k$ &           &          &          &         &           &           &           & 0.02087\footnotemark[1]   & 0.01731 & 0.01458  \\
 $l$ &           &          &          &         &           &           &           &           & 0.01643 & 0.01390  \\
 $m$ &           &          &          &         &           &           &           &           &         & 0.01327  \\
\end{tabular}
\end{ruledtabular}
\footnotetext[1] {Coincides exactly with the J-matrix result of Ref.~\cite{nasser13}.}
\end{table}
\endgroup

\section{Results and Discussion}
Before the main results are presented a few comments may be made regarding the convergence of our calculated quantities. Stability 
and accuracy of our proposed scheme is dependent mainly on the parameter $r_{\mathrm{max}}$, while apparently these are found to be 
rather insensitive with respect to variations in $\eta$ and $N$. Hence for all the calculations reported in this article, $N=300$ 
radial grid points was always found to be sufficient and accordingly also employed. Thus there is no burden on computational cost, 
as matrix sizes do not grow. Generally speaking, $r_{\mathrm{max}}=200$ or 300 a.u. was sufficiently good only for
weaker screening; in the intermediate region, an $r_{\mathrm{max}}$ of 1000 a.u. was necessary to achieve reasonable convergence. However, 
better convergence required the same to be about 1500 a.u. or so. For stronger coupling parameters as studied in this work, we had
to employ even higher values of $r_{\mathrm{max}}$ (like 7500 or so for 2p state). As expected, for higher states, while reasonable 
convergence could be achieved for similar $r_{\mathrm{max}}$ in weaker screening region, in the domain of strong coupling, even larger 
values, such as 9000 a.u. 
was employed for satisfactory convergence. Similar findings were observed when Hulth\'en and Yukawa potentials were studied within the GPS
method \cite{roy05a}. 

\begingroup
\squeezetable
\begin{table}
\caption {\label{tab:table2} Comparison of calculated negative eigenvalues (in a.u.) of MR potential for some selected
low-lying $\ell \neq 0$ states, with $\alpha=0.75$. PR signifies Present Result. See text for details.}
\begin{ruledtabular}
\begin{tabular}{l|lll}
State  & $1/b$ &  $-$E (PR)   & $-$E (Literature)          \\
\hline
2p    & 0.01   &  0.1281749227  &       \\
(0.384) & 0.025  &  0.1205273089  & 0.1205793\footnotemark[1],0.1205273\footnotemark[2]$^,$\footnotemark[6],
                                    0.1205279\footnotemark[3]$^,$\footnotemark[4],0.1205297\footnotemark[5],0.1205271\footnotemark[7]  \\
        & 0.05   &  0.1082151728  & 0.1084228\footnotemark[1],0.1082145\footnotemark[2],0.1082232\footnotemark[3],0.1082170\footnotemark[4],
                                    0.1082245\footnotemark[5],0.1082151\footnotemark[6]$^,$\footnotemark[7]  \\
        & 0.1    &  0.0852253215  & 0.0852253\footnotemark[6],0.08522531\footnotemark[8]  \\ 
        & 0.2    &  0.0459134065  & 0.04591340\footnotemark[8]              \\
        & 0.38   &  0.0004994245  & 0.000499\footnotemark[8]             \\
3p    & 0.025  &  0.0458778846  & 0.0459297\footnotemark[1],0.0458776\footnotemark[2],0.0458801\footnotemark[3],0.0458783\footnotemark[4],
                                    0.0458800\footnotemark[5],0.0458779\footnotemark[6]$^,$\footnotemark[7] \\
(0.190) & 0.05   &  0.0350633277  & 0.0352672\footnotemark[1],0.0350589\footnotemark[2],0.0350717\footnotemark[3],0.0350614\footnotemark[4],
                                    0.0350689\footnotemark[5],0.0350633\footnotemark[6]$^,$\footnotemark[7]   \\
        & 0.1    &  0.0174040195  & 0.0174040\footnotemark[6]$^,$\footnotemark[8]            \\ 
        & 0.19   &  0.0000245915  &              \\
5p    & 0.025  &  0.0098079253  & 0.0098576\footnotemark[1],0.0098055\footnotemark[2],0.0098090\footnotemark[3],0.0098062\footnotemark[4],
                                    0.0098080\footnotemark[5],0.0098079\footnotemark[6]$^,$\footnotemark[7]   \\
(0.074) & 0.072  &  0.0000854686  & 0.0000854\footnotemark[8]            \\
3d    & 0.025  &  0.0447742874  & 0.0449299\footnotemark[1],0.0447737\footnotemark[2],0.0447810\footnotemark[3],0.0447756\footnotemark[4],
                                    0.0447812\footnotemark[5],0.0447743\footnotemark[6]$^,$\footnotemark[7] \\
(0.159) & 0.05   &  0.0336929996  & 0.0343082\footnotemark[1],0.0336832\footnotemark[2],0.0337217\footnotemark[3],0.0336909\footnotemark[4],
                                    0.0337133\footnotemark[5],0.0336930\footnotemark[6]$^,$\footnotemark[7]   \\
        & 0.1    &  0.0150288223  & 0.0150288\footnotemark[6]$^,$\footnotemark[8]             \\
        & 0.158  &  0.0001124797  &             \\
4d    & 0.01   &  0.0269651708  &               \\
(0.098) & 0.025  &  0.0203017276  & 0.0204555\footnotemark[1],0.0202993\footnotemark[2],0.0203087\footnotemark[3],0.0203012\footnotemark[4],
                                    0.0208112\footnotemark[5],0.0203017\footnotemark[6]$^,$\footnotemark[7]   \\
        & 0.05   &  0.0109904267  & 0.0115742\footnotemark[1],0.0109492\footnotemark[2],0.0109919\footnotemark[3],0.0109569\footnotemark[4], 
                                    0.0109792\footnotemark[5],0.0109904\footnotemark[6]$^,$\footnotemark[7]  \\
        & 0.098  &  0.0000303175  &             \\ 
6d    & 0.025  &  0.0041649733  & 0.0043061\footnotemark[1],0.0041499\footnotemark[2],0.0041607\footnotemark[3],0.0041518\footnotemark[4], 
                                    0.0041574\footnotemark[5],0.0041650\footnotemark[6]$^,$\footnotemark[7]  \\
(0.048) & 0.047  &  0.0000655441  &             \\
4f    & 0.05   &  0.0102392570  & 0.0114284\footnotemark[1],0.0101784\footnotemark[2],0.0102639\footnotemark[3],0.0101938\footnotemark[4], 
                                    0.0102384\footnotemark[5],0.0102393\footnotemark[6]$^,$\footnotemark[7]  \\
(0.087) & 0.086  &  0.0001312291  &             \\
6f    & 0.025  &  0.0039802669  & 0.0042652\footnotemark[1],0.0039528\footnotemark[2],0.0039745\footnotemark[3],0.0039566\footnotemark[4], 
                                    0.0039677\footnotemark[5],0.0039803\footnotemark[6]$^,$\footnotemark[7]   \\
(0.044) & 0.044  &  0.0000144343  &             \\
5g    & 0.025  &  0.0090330290  & 0.0095398\footnotemark[1],0.0090190\footnotemark[2],0.0090534\footnotemark[3],0.0090254\footnotemark[4], 
                                    0.0090440\footnotemark[5],0.0090330\footnotemark[6]$^,$\footnotemark[7]  \\
(0.055) & 0.05   &  0.0010448629  & 0.0010449\footnotemark[6] \\
        & 0.054  &  0.0001267027  &             \\ 
6g    & 0.025  &  0.0037611860  & 0.0042428\footnotemark[1],0.0037220\footnotemark[2],0.0037582\footnotemark[3],0.0037284\footnotemark[4], 
                                    0.0037470\footnotemark[5],0.0037612\footnotemark[6],0.0037611\footnotemark[7] \\
(0.041) & 0.04   &  0.0001065569  &             \\ 
6h    & 0.025  &  0.0034924097  \\
(0.038) & 0.037  &  0.0001202476               \\
\end{tabular}
\end{ruledtabular}
\begin{tabbing}
$^{\mathrm{a}}${Ref.~\cite{qiang07}.}  \hspace{15pt} \= 
$^{\mathrm{b}}${Ref.~\cite{chen09}.}  \hspace{15pt} \= 
$^{\mathrm{c}}${Ref.~\cite{qiang09}.}  \hspace{15pt} \= 
$^{\mathrm{d}}${Ref.~\cite{ikhdair11}.}  \hspace{15pt} \= 
$^{\mathrm{e}}${Ref.~\cite{diaf11}.}  \hspace{15pt} \= 
$^{\mathrm{f}}${Ref.~\cite{hady11}.}  \hspace{15pt} \= 
$^{\mathrm{g}}${Ref.~\cite{lucha99}.}  \hspace{15pt} \= 
$^{\mathrm{h}}${Ref.~\cite{nasser13}.}  
\end{tabbing}
\end{table}
\endgroup

Now in Table I, we report the estimated critical screening parameters $(b^{-1})_c$ of MR potential for all the 55 eigenstates 1s through 10m, 
having $\alpha=0.75$. For a particular bound state, this is defined as the value of $1/b$, beyond which the state does not appear in the 
bound-state spectrum. Alternatively, this corresponds to a value of the parameter at which energy of such a state is zero. These are
important quantities as they play a major role in limiting the accuracy of calculated results. These have been well studied by a number of 
authors for two common screened Coulomb potentials, such as Hulth\'en \cite{varshni90} and Yukawa \cite{rogers70} potentials. However, for 
MR potential, the only such attempt has been made in a J-matrix calculation \cite{nasser13} for the non-zero rotational states having 
vibrational quantum number 
$n=2-8$. As seen from the table, our results completely agree with those of J-matrix result \cite{nasser13}. For $s$-waves and 
$n=9,10,$ they are reported here for the first time. In general, for a given $\alpha$, $(b^{-1})_c$ tends to decrease with increase in 
$n$ and $\ell$ quantum numbers. A similar exercise was done for for all the 55 states ($1 \leq n \leq 10$) of $\alpha=1.5$ as well. While no 
attempt is made to do a systematic
study of the effect of $\alpha$ on $(b^{-1})_c$, it is noticed that, as $\alpha$ goes from 0.75 to 1.5, critical screening parameter for a given 
state decreases. These critical values for $\alpha=1.5$ are not produced here, as they do not add any further insight into our understanding. 
Instead, individual values are supplied in future tables, as needed.

\begingroup
\squeezetable
\begin{table}
\caption {\label{tab:table3} Comparison of calculated negative eigenvalues (in a.u.) of MR potential for some selected
low-lying $\ell \neq 0$ states, with $\alpha=1.5$. PR signifies Present Result. See text for details.} 
\begin{ruledtabular}
\begin{tabular}{l|lll}
State  & $1/b$ &  $-$E (PR)   & $-$E (Literature)   \\
\hline
2p        & 0.01   & 0.0961495847   &  \\       
(0.352)     & 0.025  & 0.0899708754   & 0.0900229\footnotemark[1],0.0899708\footnotemark[2]$^,$\footnotemark[7],0.0899721\footnotemark[3],
                                 0.0899715\footnotemark[4],0.0899732\footnotemark[5],0.0899709\footnotemark[6]  \\
     & 0.04   & 0.0839572012   & \\  
     & 0.05   & 0.0800399908   & 0.0802472\footnotemark[1],0.0800389\footnotemark[2],0.0800492\footnotemark[3],0.0800414\footnotemark[4], 
                                 0.0800489\footnotemark[5],0.0800400\footnotemark[6]$^,$\footnotemark[7] \\
     & 0.2    & 0.0303447183   & \\
     & 0.35   & 0.0002422083   & \\
3p & 0.025  & 0.0369133922   & 0.0369651\footnotemark[1],0.0369130\footnotemark[2],0.0369157\footnotemark[3],0.0369137\footnotemark[4], 
                                 0.0369154\footnotemark[5],0.0369134\footnotemark[6]$^,$\footnotemark[7] \\ 
(0.173)  & 0.05   & 0.0272696509   & 0.0274719\footnotemark[1],0.0272636\footnotemark[2],0.0272769\footnotemark[3],0.0272662\footnotemark[4], 
                                 0.0272736\footnotemark[5],0.0272697\footnotemark[6],0.0272696\footnotemark[7]  \\
     & 0.1    & 0.0119726070   & \\       
     & 0.17   & 0.0002084981   & \\
5p & 0.025  & 0.0080816394   & 0.0081308\footnotemark[1],0.0080787\footnotemark[2],0.0080822\footnotemark[3],0.0080793\footnotemark[4], 
     0.0080812\footnotemark[5],0.0080816\footnotemark[6]$^,$\footnotemark[7] \\
(0.068)  & 0.068  & 0.0000169949   & \\
3d & 0.025  & 0.0394789425   & 0.0396345\footnotemark[1],0.0394782\footnotemark[2],0.0394860\footnotemark[3],0.0394801\footnotemark[4], 
                                 0.0394857\footnotemark[5],0.0394789\footnotemark[6]$^,$\footnotemark[7] \\  
(0.153) & 0.05   & 0.0294495639   & 0.0300629\footnotemark[1],0.0294379\footnotemark[2],0.0294773\footnotemark[3],0.0294456\footnotemark[4], 
                                 0.0294680\footnotemark[5],0.0294496\footnotemark[6]$^,$\footnotemark[7]  \\
     & 0.1    & 0.0125825188   &  \\      
     & 0.15   & 0.0006473604   &   \\
4d & 0.01   & 0.0244810816   &    \\    
(0.094) & 0.025  & 0.0182114637   & 0.0183649\footnotemark[1],0.0182087\footnotemark[2],0.0182182\footnotemark[3],0.0182106\footnotemark[4], 
                                 0.0182162\footnotemark[5],0.0182115\footnotemark[6]$^,$\footnotemark[7]  \\
     & 0.05   & 0.0095166719   & 0.0100947\footnotemark[1],0.0094967\footnotemark[2],0.0095129\footnotemark[3],0.0094775\footnotemark[4], 
                                 0.0094998\footnotemark[5],0.0095167\footnotemark[6]$^,$\footnotemark[7]  \\ 
     & 0.094  & 0.0000908452   &     \\ 
6d & 0.025  & 0.0036813104   & 0.0038209\footnotemark[1],0.0036647\footnotemark[2],0.0036756\footnotemark[3],0.0036666\footnotemark[4], 
                                 0.0036722\footnotemark[5],0.0036813\footnotemark[6]$^,$\footnotemark[7] \\
(0.046) & 0.046  & 0.0000235387   &          \\
4f & 0.05   & 0.0094014592   & 0.0105852\footnotemark[1],0.0093353\footnotemark[2],0.0094212\footnotemark[3],0.0093507\footnotemark[4], 
                                 0.0093953\footnotemark[5],0.0094015\footnotemark[6]$^,$\footnotemark[7] \\
(0.085)   & 0.085  & 0.0000705882   &          \\
6f & 0.025  & 0.0036774476   & 0.0039606\footnotemark[1],0.0036481\footnotemark[2],0.0036699\footnotemark[3],0.0036520\footnotemark[4], 
                                 0.0036631\footnotemark[5],0.0036774\footnotemark[6]$^,$\footnotemark[7] \\
(0.043)  & 0.043  & 0.0000339418   &          \\
5g & 0.025  & 0.0086150371   & 0.0091210\footnotemark[1],0.0086002\footnotemark[2],0.0086347\footnotemark[3],0.0086066\footnotemark[4], 
                                 0.0086252\footnotemark[5],0.0086150\footnotemark[6]$^,$\footnotemark[7] \\
(0.054) & 0.05   & 0.0009055496   & 0.0009055\footnotemark[6]         \\
     & 0.053  & 0.0002324402   &          \\ 
6g & 0.025  & 0.0035623305   & 0.0040422\footnotemark[1],0.0035214\footnotemark[2],0.0035576\footnotemark[3],0.0035278\footnotemark[4], 
                                 0.0035464\footnotemark[5],0.0035623\footnotemark[6]$^,$\footnotemark[7] \\
(0.040)  & 0.04   & 0.0000378830   &   \\
6h & 0.025  & 0.0033607331   &    \\
(0.037) & 0.037  & 0.0000707576   &          \\
\end{tabular}
\end{ruledtabular}
\begin{tabbing}
$^{\mathrm{a}}${Ref.~\cite{qiang07}.}  \hspace{25pt} \= 
$^{\mathrm{b}}${Ref.~\cite{chen09}.}  \hspace{25pt} \= 
$^{\mathrm{c}}${Ref.~\cite{qiang09}.}  \hspace{25pt} \= 
$^{\mathrm{d}}${Ref.~\cite{ikhdair11}.}  \hspace{25pt} \=  
$^{\mathrm{e}}${Ref.~\cite{diaf11}.}  \hspace{25pt} \= 
$^{\mathrm{f}}${Ref.~\cite{hady11}.}  \hspace{25pt} \= 
$^{\mathrm{g}}${Ref.~\cite{lucha99}.}  
\end{tabbing}
\end{table}
\endgroup

Now in Tables II and III, our calculated ro-vibrational energies are given for selected $\ell \neq 0$ states having vibrational 
quantum number, $n \leq 6$, for $\alpha=0.75$ and 1.5 respectively. In both cases, large range of screening parameters, covering weak, 
medium and strong interaction are considered. Numbers in the parentheses in Column 1 denote the truncated values of critical 
screening parameters as estimated in \cite{nasser13}, and also independently confirmed in present work. These values do not exist for 
$\alpha=1.5$ in the literature. In several occasions, many reference eigenvalues exist for both $\alpha$; some of these are quoted here 
for comparison. Note that, generally there is a predominance of excellent quality 
results in low screening parameter region and scarcity of same for larger screening; here we have focused more on the latter. 
Thus, we have gone \emph{beyond the interaction region} considered in any of the previous works so far, for practically all the states in these tables.
Reference energies of \cite{qiang07}, obtained through an approximation of $\frac{1}{r^2}$ in terms of the generalized 
hyper-geometric functions $_2F_1(a,b;c;z)$, are reasonably good for small $\alpha$. However, it performs rather less accurately for
higher vibrational and rotational quantum number, as well as for screening parameters in the neighborhood of $(b^{-1})_c$. Some of these
states are also reported from super-symmetric shape invariance approach and wave function analysis \cite{chen09} along with a constant-introduced
new approximation for centrifugal term. Quality of the energies in this case is slightly better than the previous case \cite{qiang07}; 
however shows 
a similar pattern of discrepancies widening for higher $n,\ell$ and $(b^{-1})$, as earlier. Overall similar quality eigenvalues are also obtained 
for 2p--6g states for both $\alpha$ values in \cite{qiang09}, via an approximation for centrifugal term containing three adjustable parameters. 
Good-quality energies (better than \cite{chen09}) are also reported in \cite{ikhdair11} within the rubric of Nikiforov-Uvarov method and 
employing yet another new approximation for centrifugal term. Approximate analytic energies were obtained maintaining, more or less, a 
similar accuracy pattern as in all the methods mentioned above. Moderate quality results and similar trend in energy behavior have been reported 
from Feynman path integral formalism in conjunction with an improved $\frac{1}{r^2}$ approximation \cite{diaf11}. However, it seems that, 
so far, the
best energies are those given in \cite{hady11}, where the reference Hamiltonian is tridiagonalized in Laguerre and oscillator bases, 
and MR potential matrix elements are calculated using Gauss quadrature approach. Excepting very slight deviations (0.0000001 a.u.) in only 
two cases (6g state for $\alpha=0.75$, $1/b=0.025$; and 3p state with $\alpha=1.5$, $1/b=0.05$), these energies completely coincide with 
the numerical estimates obtained from MATHEMATICA \cite{lucha99}. Our results, reported here with much better precision, show excellent 
agreement with these energies for all states. 
As already mentioned, in most cases, not enough results are available in larger screening region. So such states 
presented in this work can not be directly compared. For $\alpha=0.75$, only some selected states have been lately published through a J-matrix
method \cite{nasser13} in high screening region; present eigenvalues compare excellently with these values in all occasions. No result
could be found for 6h states. These and some others are reported here for the first time. 
Note that, all our energies are reported for more decimal places than the existing methods in literature. Considering the performance
of GPS method for various physical systems in past years, we believe that current energies may be the most accurate estimates 
reported so far, \emph{surpassing} all the reference works mentioned above. 

\begingroup
\squeezetable
\begin{table}
\caption {\label{tab:table4} Calculated negative eigenvalues (a.u.) of MR
potential for $\ell \neq 0, n=8,10$ states at $\alpha$ of 0.75. 
Numbers in the parentheses denote critical screening parameters.} 
\begin{ruledtabular}
\begin{tabular}{llllll}
 State & $1/b$ & $-$Energy & State &  $1/b$ & $-$Energy   \\ \hline 
8p (0.030) & 0.02 & 0.0010509525    & 10p (0.019) & 0.01 & 0.0012865066    \\
8d (0.028) & 0.02 & 0.0009679472    & 10d (0.019) & 0.01 & 0.0012520358    \\
8f (0.027) & 0.02 & 0.0008758189    & 10f (0.018) & 0.01 & 0.0012218240    \\
8g (0.025) & 0.02 & 0.0007601651    & 10g (0.017) & 0.01 & 0.0011877190    \\
8h (0.024) & 0.02 & 0.0006146565    & 10h (0.016) & 0.01 & 0.0011473727    \\
8i (0.022) & 0.02 & 0.0004330625    & 10i (0.015) & 0.01 & 0.0010996172    \\
8k (0.021) & 0.02 & 0.0002062354    & 10k (0.015) & 0.01 & 0.0010435741    \\
            &      &                  & 10l (0.014) & 0.01 & 0.0009783672    \\
            &      &                  & 10m (0.013) & 0.01 & 0.0009029721    \\  
\end{tabular}
\end{ruledtabular}
\end{table}
\endgroup

After low-lying states, as a testimony of the usefulness of GPS approach, representative energies are now offered for some 
higher states, which are quite scarce in literature. Table IV thus tabulates all the $\ell \neq 0$ eigenstates for $n=8,10$ at selected value of 
$\alpha=0.75$ having $1/b=0.02$ and 0.01 respectively. 
In the parentheses, calculated critical screening values after truncation, are once again quoted for convenience. For $n=8$, these are also
available from \cite{nasser13}, as discussed in Table I. As $n$, $\ell$ increase, calculation of these states
become progressively difficult. Thus very few attempts beyond 6g have been recorded so far; \emph{viz.}, (i) 8p--8f states for both 
$\alpha=0.75,1.5$ having $1/b=0.025$ were studied by means of Laguerre and oscillator bases \cite{hady11} (ii) 8p state for 
$1/b=0.025, 0.029$, and 8k for $1/b=0.019$, all for $\alpha=0.75$, via the J-matrix method \cite{nasser13}. However, for $n=10$, we are not
aware of any results. In our test calculation, these results were reproduced quite nicely. Hence these states are given here for first time
and can not be directly compared with reference values in the literature. It is hoped that they may be helpful for future referencing. 

\begin{figure}
\begin{minipage}[c]{0.40\textwidth}
\centering
\includegraphics[scale=0.45]{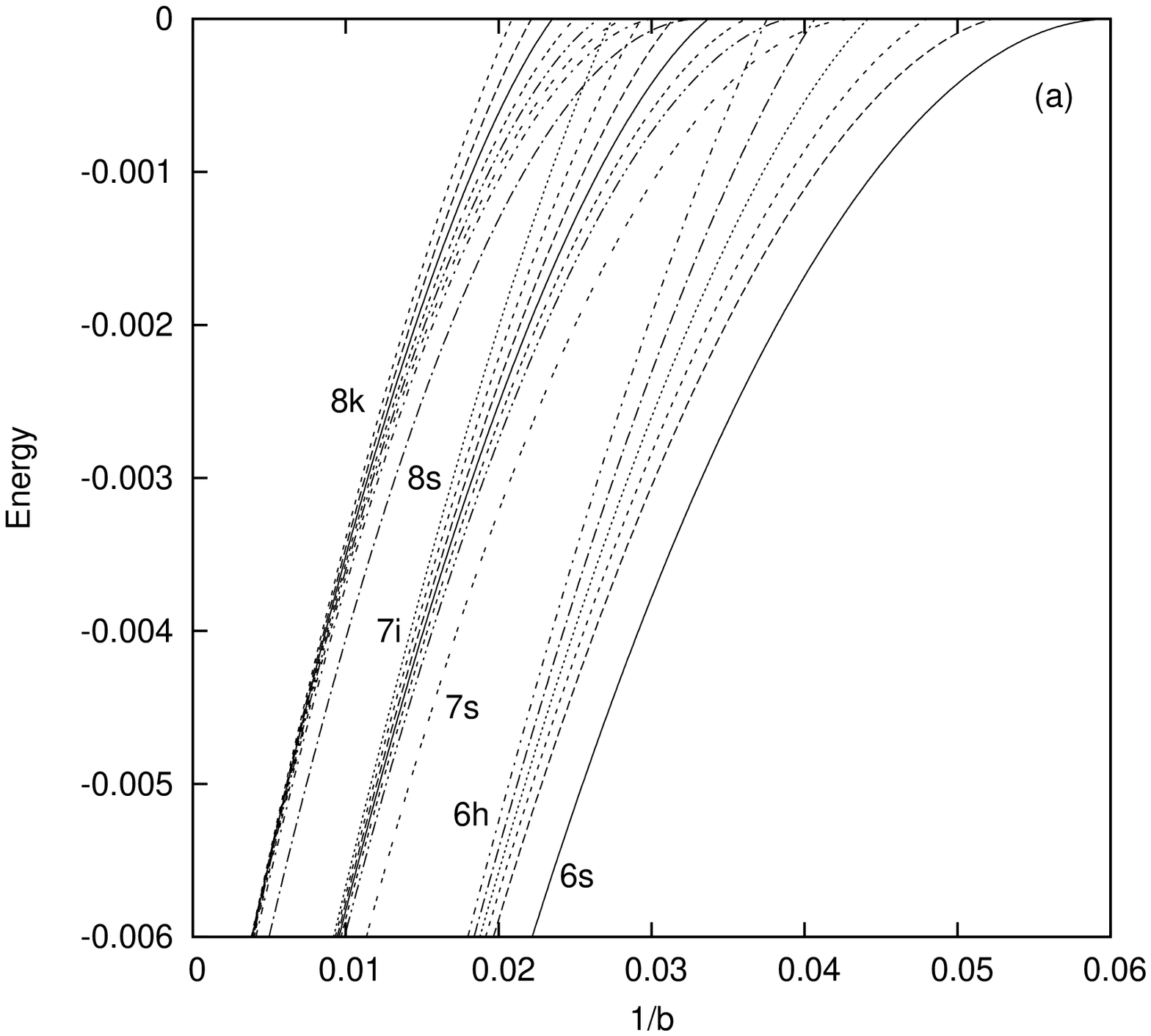}
\end{minipage}%
\hspace{0.5in}
\begin{minipage}[c]{0.40\textwidth}
\centering
\includegraphics[scale=0.45]{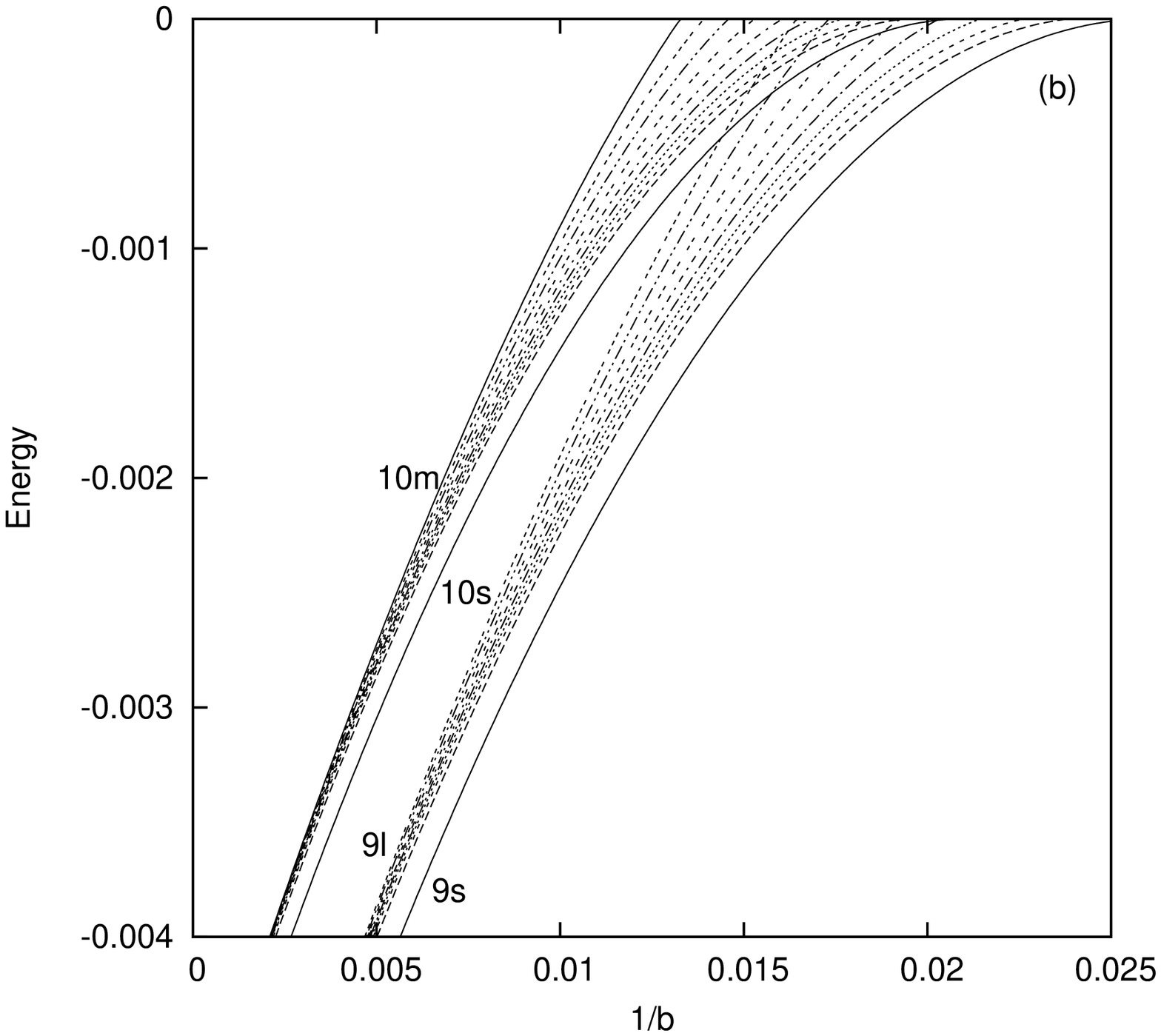}
\end{minipage}%
\caption{Energy eigenvalues (a.u.) of MR potential for (a) $n=6,7,8$ and (b) $n=9,10$ levels 
respectively, as function of $1/b$ in the vicinity of zero energy. In both cases, $\alpha=0.75$.}
\end{figure}

Next Fig.~1 depicts the variation of energy eigenvalues as function of $1/b$ for all the states of MR potential belonging to  $n=6,7,8$ 
(left) and $n=9,10$ (right) respectively; all having $\alpha=0.75$. For a given $\alpha$, energies tend to increase with an increase in 
screening. Generally, while the individual non-zero $\ell$-states remain very closely spaced together for a chosen $n$, 
$s$-waves slightly separate them from others, with progressive lowering of separation for higher $n$. However, the states of a given $n$ level 
generally remain well separated from other $n$ levels, 
for small values of vibrational quantum number. But with an increase in $n$, significant deviations from such simple unmixed ordering is 
encountered leading to complex level crossing in the neighborhood of zero energy. This probability of mixing gradually increases with $n$ 
and for higher $n$, accurate evaluation of such energy levels becomes rather difficult due to the heavy mixing amongst these states. Thus, 
in the left-hand side (a) slight mixing is 
observed among 6g, 6h and 7s, 7p at around $1/b=0.037-0.040$; again 7g, 7h, 7i mixing with 8s, 8p, 8d with some greater intensity at around
$1/b=0.025-0.030$. However, in the right-hand side (b), the mixing is much more pronounced among 9i, 9k, 9l and 10s, 10p, 10d, 10f at around 
$1/b=0.017-0.020$. One also notices that for a given value of $n$, the separation between states having different $\ell$ steadily increases 
with an increase in $1/b$. This is reminiscent of the energy orderings in Hulth\'en and Yukawa potentials \cite{varshni90, roy05a}.   

\begin{figure}
\begin{minipage}[c]{0.40\textwidth}
\centering
\includegraphics[scale=0.45]{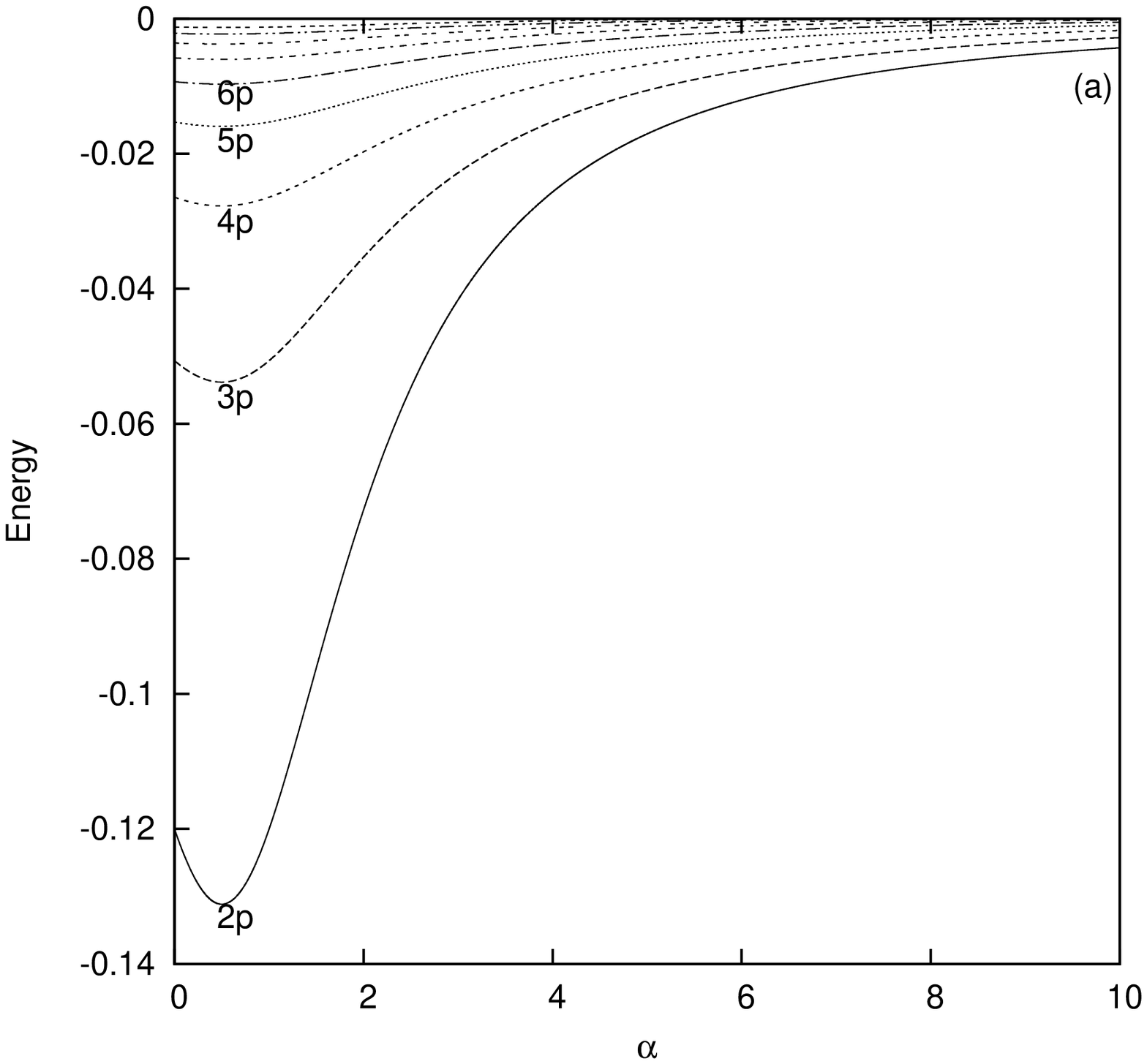}
\end{minipage}%
\hspace{0.5in}
\begin{minipage}[c]{0.40\textwidth}
\centering
\includegraphics[scale=0.45]{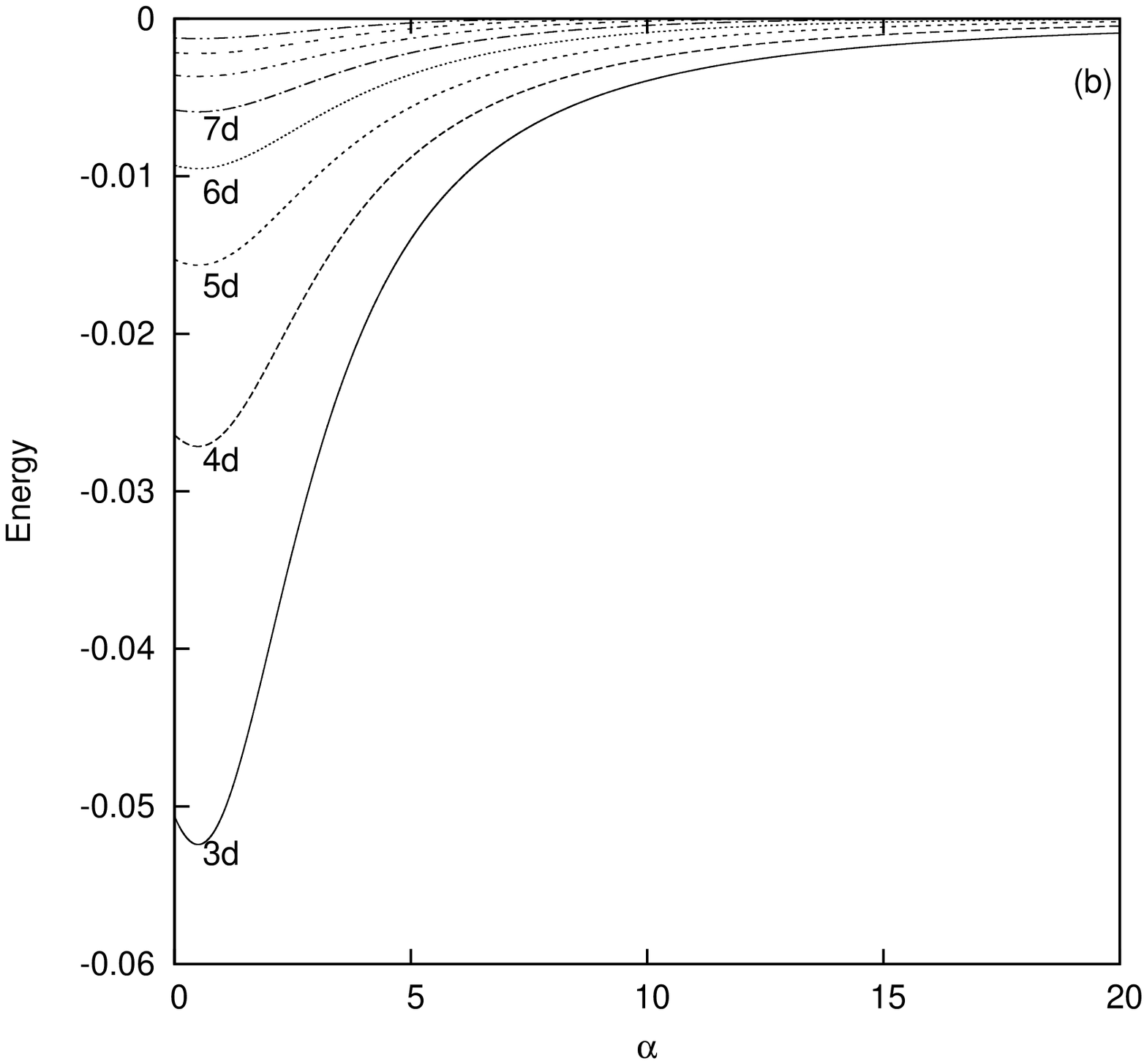}
\end{minipage}%
\caption{Energy eigenvalues (a.u.) of MR potential for (a) np, $n=2-10$ and (b) nd, $n=3-10$ levels 
respectively, as function of $\alpha$, at a fixed value of $1/b=0.01$.}
\end{figure}

Next, energy variations of MR potential with respect to changes in $\alpha$ are discussed for representative states. For this Fig.~2
displays the behavior of (a) np and (b) $n'$d states ($n= 2-10$, $n'=3-10$) at a screening corresponding to $1/b=0.01$. Note that
the maximum in $\alpha$ axis is different for these two plots. All these states follow similar qualitative pattern; in the beginning, 
at smaller region of $\alpha$, any increase causes a lowering in energy until it reaches a minimum, followed by a sharp increase, and finally 
tends towards zero slowly. A similar trend in behavior was observed for other states as well. For a given value of rotational quantum number, 
as $n$ assumes higher values, the well becomes progressively
shallow and flatter, so much so that for higher states like 10p or 10d, the plots nearly approach a straight line. Also for smaller 
$n$, the individual $\ell$ states remain distinct and well separated; however, as $n$ takes higher values, separation between the members 
narrows down making them very closely spaced to each other.

\begin{figure}
\begin{minipage}[c]{0.40\textwidth}
\centering
\includegraphics[scale=0.45]{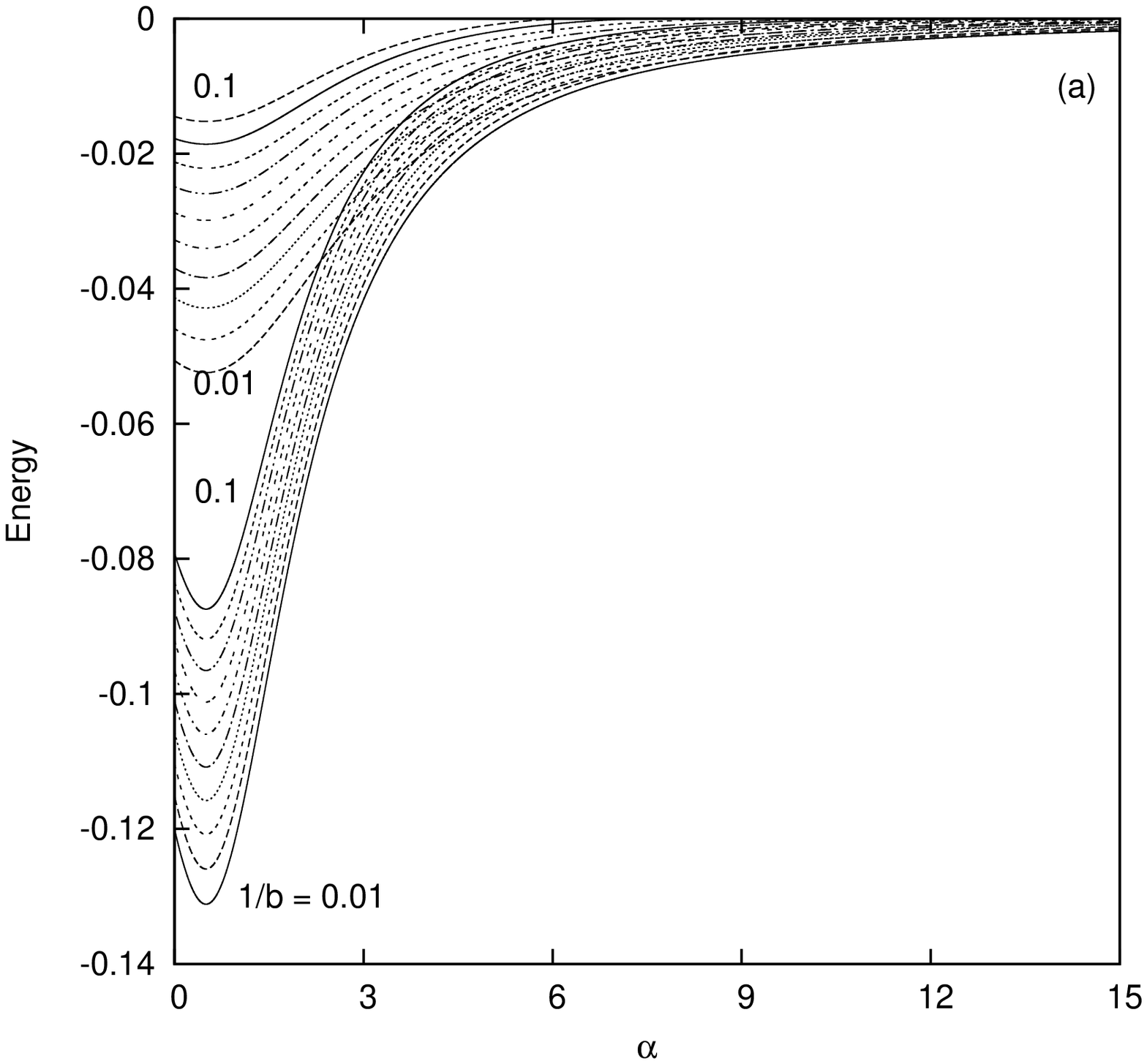}
\end{minipage}%
\hspace{0.5in}
\begin{minipage}[c]{0.40\textwidth}
\centering
\includegraphics[scale=0.45]{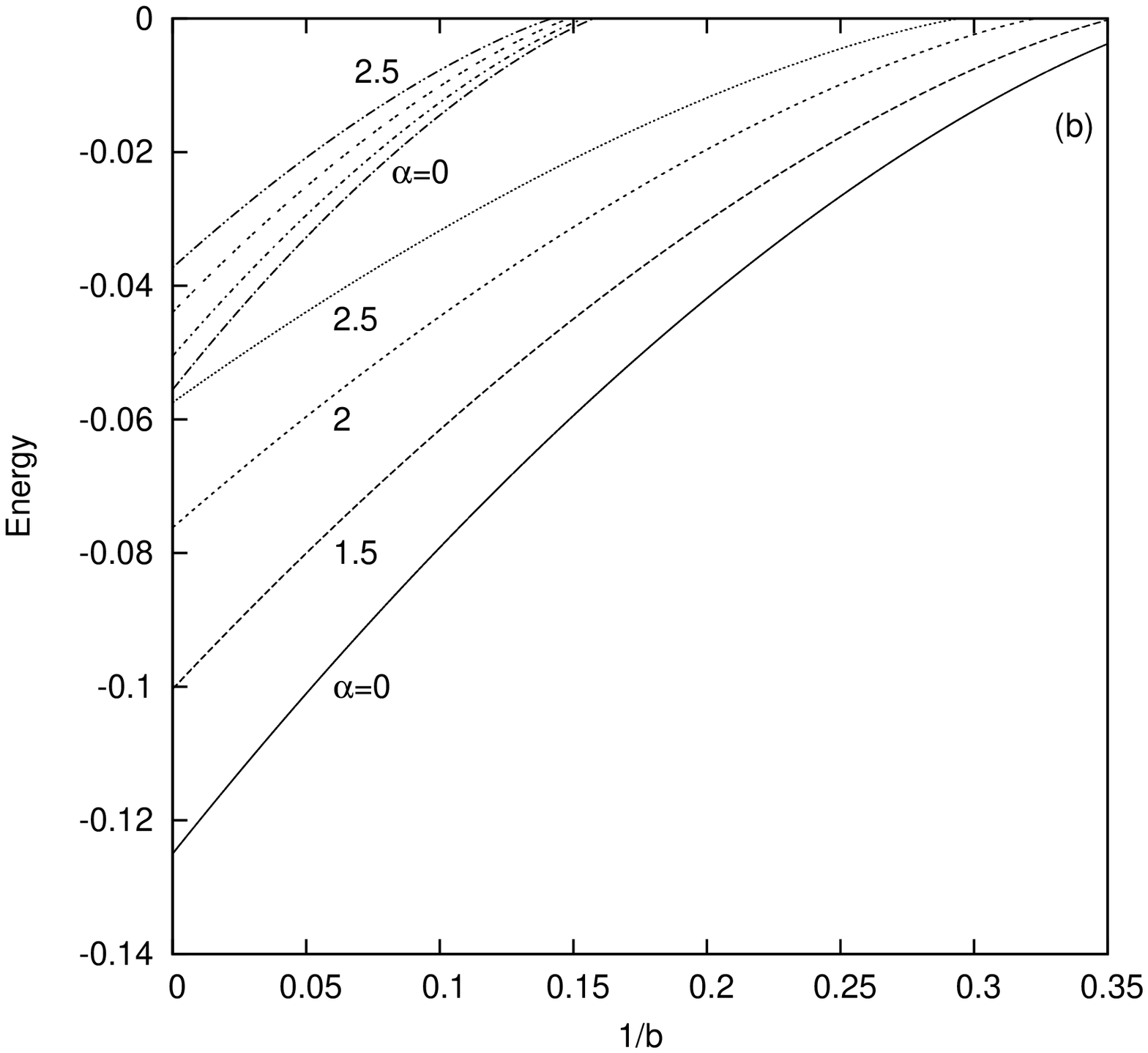}
\end{minipage}%
\caption{Energy eigenvalues (a.u.) of MR potential as function of (a) $\alpha$, for $1/b=0.01,02,0.03,\cdots ,0.10$, and 
(b) $1/b$, for fixed values of $\alpha =0,1.5,2,2.5,$ respectively. Bottom and top segments correspond to 2p and 3d (top) states
respectively.}
\end{figure}

Figure 3 displays the energy changes of MR potential as functions of $\alpha$ in left (a) for ten selected $1/b$ values \emph{viz.}, 
0.01, 0.02, 0.03, $\cdots$, 0.1, covering a broad range of interaction. In (b) likewise is shown the variations with respect to $1/b$, for
4 representative $\alpha$ values, namely 0, 1.5, 2, 2.5. In both these figures, bottom and top family of plots correspond to 2p and 3d 
states respectively. Not surprisingly, in (a) and (b), we see the trend as expected from Fig.~2 and Fig.~1 respectively. With an 
increase in the screening parameter, minimum in energy is gradually shifted to higher values in (a). Once again in (a), the plots are
much flatter for 3d compared to those of 2p. One also notices appreciable mixing of 2p and 3d levels starting approximately
at $\alpha =3$ which continues thereafter more vigorously, as the energies approach towards zero. In (b), 2p, 3d levels are well separated; 
however the gap decreases in 3d with an increase in $\alpha$ value. Energies increase with an increase in $1/b$ for a particular $\alpha$.  
We also examined the nature of such plots for large $n$ states, which are not shown here for brevity. One generally finds that, for a given 
screening parameter, similar plots in (a) tend to become flatter as we proceed towards higher values of $n$, within a given $\ell$, so much so 
that for 10p, or 10d, they very closely resemble a straight line.

\begin{figure}
\begin{minipage}[c]{0.40\textwidth}
\centering
\includegraphics[scale=0.45]{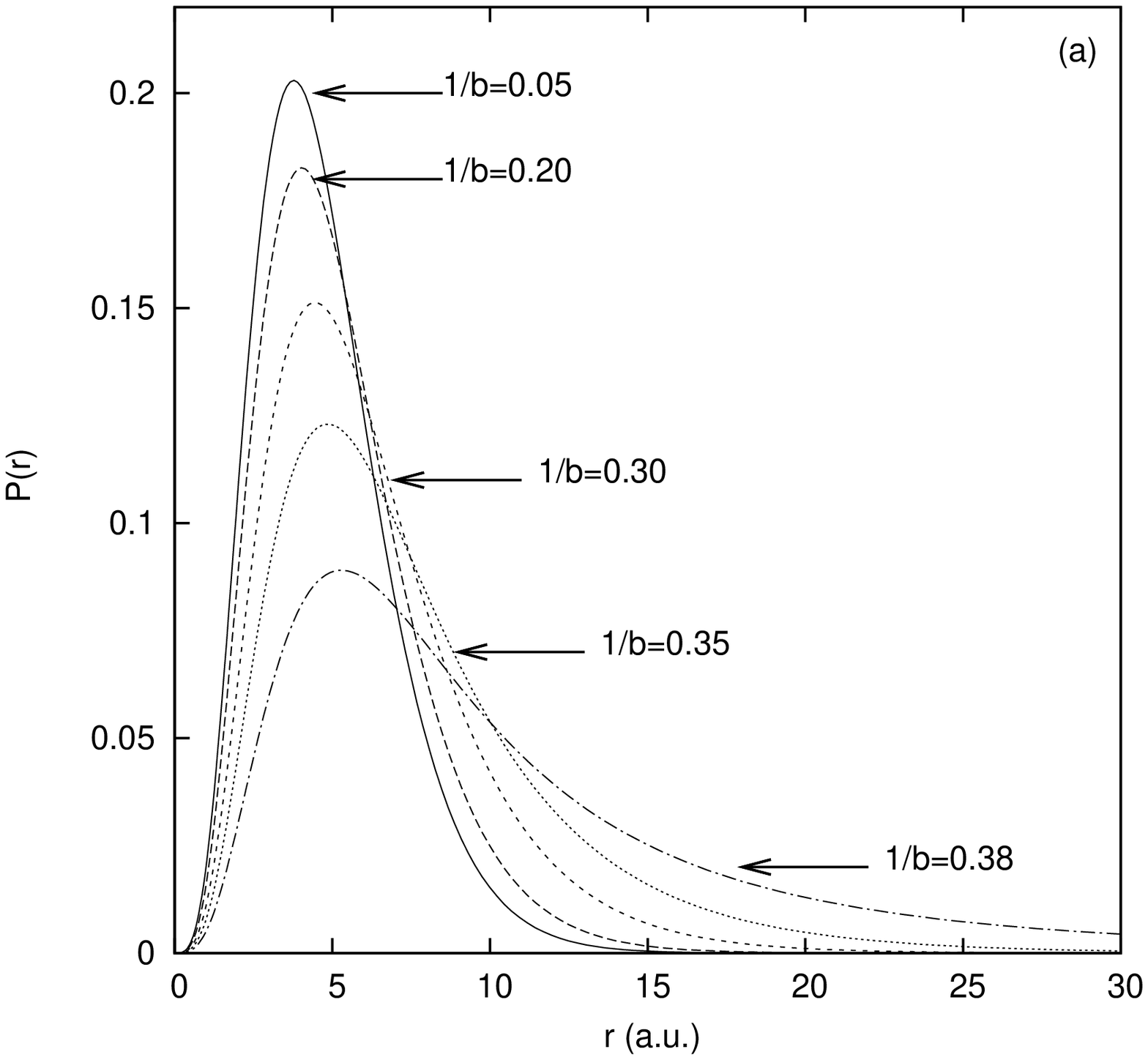}
\end{minipage}%
\hspace{0.5in}
\begin{minipage}[c]{0.40\textwidth}
\centering
\includegraphics[scale=0.45]{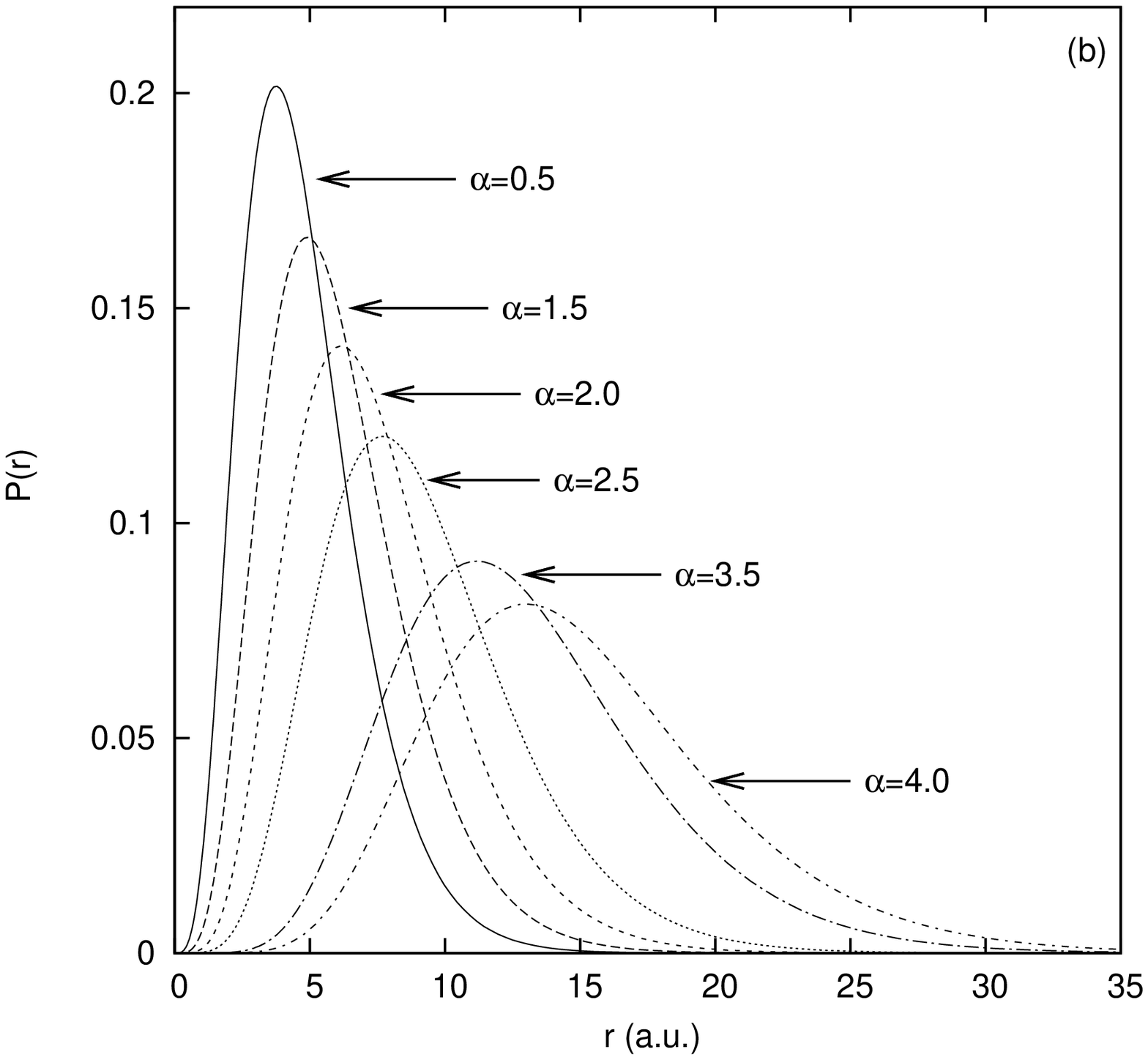}
\end{minipage}%
\caption{Radial probability distribution functions, P(r) (in a.u.), of 2p state of MR potential for (a) fixed $\alpha=0.75$, at 
five $1/b$ values and (b) fixed $1/b=0.1$, at six $\alpha$ values, as indicated in figure.}
\end{figure}

The left portion (a) of Fig.~4 now depicts the characteristic features of radial probability distribution function for 2p state of MR 
potential at a chosen value
of $\alpha=0.75$, for five screening parameters, covering weak ($1/b=0.05$)-intermediate ($1/b=0.2,0.3$)-strong ($1/b=0.35, 0.38$) interaction.  
It is seen that with an increase in screening parameter, the density distribution oozes out to progressively larger values of $r$, while peak 
values are reduced sharply. The peak positions, however, do not show any visible shifting. The deep, narrow curves in low screening region
spread quite 
significantly with increasing screening effect bringing flatness into the picture. Consequently, tail of the wave functions extend to much 
larger $r$ in the latter scenario. This may be partly responsible for larger values of $r_{\mathrm{max}}$ needed in our calculation in  
strong screening regions. Similar phenomenon has been observed in
J-matrix calculation \cite{nasser13}. In the right segment (b), similar distribution functions of MR potential are displayed for 2p state, at 
a fixed $1/b=0.1$. Six $\alpha$ values in the region of 0.5--4 are chosen for this purpose, as indicated in the figure. 
As $\alpha$ increased, the radial density spreads in substantial amount; height of the peak gradually reduced and shifted to higher values 
of $r$. Thus one may intuitively expect that relatively larger $r_{\mathrm{max}}$ would be required as one goes to higher values of $\alpha$. 

\begingroup
\squeezetable
\begin{table}
\caption {\label{tab:table5} Calculated expectation values (a.u.) of MR potential for some selected states. Left and right panels 
correspond to $\alpha=0.75$, 1.5 respectively. $1/b$ is 0.1 in all cases.} 
\begin{ruledtabular}
\begin{tabular}{l|llll|llll}
State  & \multicolumn{4}{c}{$\alpha=0.75, 1/b=0.1$}      &  \multicolumn{4}{c}{$\alpha=1.5, 1/b=0.1$}  \\
\cline{2-9}
       & $\langle r^{-2} \rangle $ & $\langle r^{-1} \rangle $ & $\langle r \rangle$ & $ \langle r^2  \rangle$  
       & $\langle r^{-2} \rangle $ & $\langle r^{-1} \rangle $ & $\langle r \rangle$ & $ \langle r^2  \rangle$  \\ \hline 
2p   &   0.091584   &  0.260223  &  4.850846  &  28.451982  & 0.053464  & 0.203219 &  6.050858  &  43.514317   \\
       &              &  0.260208\footnotemark[1]  & 4.850753\footnotemark[1]  &     &    &    &    &          \\
3p   &   0.022569   &  0.104064  &  13.353515 & 206.302341  & 0.014398 &  0.086702 & 15.729681   &  285.71142   \\
       &              &  0.104060\footnotemark[1]   &  13.353791\footnotemark[1]   & &  &  &  & \\
4p   &   0.004549   &  0.039654  &  34.842184 & 1411.5715   & 0.002030  & 0.027594 &  50.852549  &  3234.38358   \\
       &              &  0.039636\footnotemark[1]   &  34.851618\footnotemark[1]       &  &  &  &  & \\
3d   &  0.013003    & 0.103225   &  11.489452   &  153.35102  & 0.010760   & 0.094364 & 12.487876  &  180.39909     \\
       &              &  0.103223\footnotemark[1]   &  11.488341\footnotemark[1]       &  &  &  &  & \\

\end{tabular}
\end{ruledtabular}
\footnotetext[1] {Ref.~\cite{nasser13}.}
\end{table}
\endgroup

As a further test of the quality and convergence of our eigenfunctions, four calculated density moments, \emph{viz.}, $\langle r^{-2} \rangle $,
$\langle r^{-1} \rangle $, $\langle r \rangle$, and $ \langle r^2  \rangle$, of MR potential are reported for 2p, 3p, 4p, 3d states; all having
$1/b=0.1$ and corresponding to $\alpha=0.75$ (left panel) and 1.5 (right panel). These have not been reported before except for some 
$\langle r^{-1} \rangle $ and $\langle r \rangle$, for $\alpha=0.75$, in J-matrix study of \cite{nasser13}, which are duly quoted for 
comparison. The general agreement seems to be quite satisfactory. However, some minor discrepancies are noticed, especially for 
$\langle r \rangle$; which could be due to the differences in wave functions obtained from these two methods. 

\section{conclusion}
A detailed study has been carried out on the \emph{accurate} eigenvalues, eigenfunctions, density moments and radial densities of MR potential by means
of GPS formalism. This is a simple, quite easy to implement, accurate and reliable method. \emph{All} the 55 eigenstates belonging to $n \leq 10$ 
have been presented with excellent accuracy. Results are compared wherever possible. A detailed estimate of the \emph{critical screening} 
parameter is provided and compared. A thorough analysis is made to examine the \emph{effect
of screening} by scanning through weak, intermediate and very stronger regions. Special emphasis has been given to \emph{stronger} couplings 
as there is a visible scarcity in the literature. Considerable attention was also paid for \emph{high-lying states}, as reference values are 
quite scanty for these states. As demonstrated, present results are \emph{significantly improved} from the best reference results available so 
far.  Many \emph{new} states are reported here for the first time. In view of the
simplicity and accuracy offered by this method for the system under investigation, it is hoped that this may be equally useful and successful 
for other molecular potentials, some of which may be taken up in later communications.   



\begin{thebibliography}{99}
\bibitem{manning33} M.~F.~Manning and N.~Rosen, Phys.~Rev.~ \textbf{44}, 953 (1933).
\bibitem{leroy70} R.~J.~Le Roy and R.~B.~Bernstein, J.~Chem.~Phys.~ \textbf{52}, 3869 (1970).
\bibitem{cai86} J.~Cai, P.~Cai and A.~Inomata, Phys.~Rev.~A \textbf{34}, 4621 (1986).
\bibitem{hulthen42} L.~Hulth\'en, Ark.~Mat.~Astron.~Fys.~ \textbf{28A}, 5 (1942); \textbf{29B}, 1 (1942).
\bibitem{infeld51} I.~Infeld and T.~E.~Hull, Rev.~Mod.~Phys.~ \textbf{23}, 21 (1951). 
\bibitem{diaf05} A.~Diaf, A.~Chouchaoui and R.~J.~Lombard, Ann.~Phys.~ \textbf{317}, 354 (2005).
\bibitem{dong07} S.-H.~Dong and J.~Garc\'ia-Ravelo, Phys.~Scr.~ \textbf{75}, 307 (2007). 
\bibitem{mincang10} Z.~Min-Cang and A.~Bo, Chin.~Phys.~Lett.~ \textbf{27}, 11 (2010).
\bibitem{chen07} C.-Y.~Chen, F.-L.~Lu and D.-S.~Sun, Phys.~Scr.~ \textbf{76}, 428 (2007).
\bibitem{qiang07} W.-C.~Qiang and S.-H.~Dong, Phys.~Lett.~A \textbf{368}, 13 (2007).
\bibitem{chen09} Z.-Y.~Chen, M.~Li and C.-S.~Jia, Mod.~Phys.~Lett.~A \textbf{24}, 1863 (2009).
\bibitem{qiang09} W.-C.~Qiang and S.-H.~Dong, Phys.~Scr.~ \textbf{79}, 045004 (2009).
\bibitem{ikhdair11} S.~M.~Ikhdair, Phys.~Scr.~ \textbf{83}, 015010 (2011). 
\bibitem{diaf11} A.~Diaf and A.~Chouchaoui, Phys.~Scr.~ \textbf{84}, 015004 (2011). 
\bibitem{hady11} A.~Abdel-Hady, Proceedings of the 8$^{th}$ Conference on Nuclear and Particle Physics, Hurghada, Egypt. pp.~131 (2011).
\bibitem{nasser13} I.~Nasser, M.~S.~Abdelmonem and A.~Abdel-Hady, Mol.~Phys.~ \textbf{111}, 1 (2013).
\bibitem{wei08} G.-F.~Wei, C.-Y.~ Long and S.-H.~Dong, Phys.~Lett.~A \textbf{372}, 2592 (2008). 
\bibitem{gu11} X.-Y.~Gu and S.-H.~Dong, J.~Math.~Chem.~ \textbf{49}, 2053 (2011).
\bibitem{arda12} A.~Arda and R.~Sever, J.~Math.~Chem.~ \textbf{50}, 1920 (2012).
\bibitem{lucha99} W.~Lucha and F.~F.~Sch\"oberl, Int.~J.~Mod.~Phys.~C \textbf{10}, 607 (1999). 
\bibitem{roy04} A.~K.~Roy, Phys.~Lett.~A \textbf{321}, 231 (2004).
\bibitem{roy04b} A.~K.~Roy, J.~Phys.~B \textbf{37}, 4369 (2004); {\it ibid.} \textbf{38}, 1591 (2005).
\bibitem{roy05} A.~K.~Roy, Int.~J.~Quant.~Chem.~\textbf{104}, 861 (2005).
\bibitem{roy05a} A.~K.~Roy, Pramana--J.~Phys.~ \textbf{65}, 01 (2005). 
\bibitem{roy07} A.~K.~Roy and A.~F.~Jalbout, Chem.~Phys.~Lett.~ \textbf{445}, 355 (2007). 
\bibitem{roy08} A.~K.~Roy, A.~F.~Jalbout and E.~I.~Proynov, Int.~J.~Quant.~Chem.~ \textbf{108}, 
827 (2008). 
\bibitem{roy08a} A.~K.~Roy, A.~F.~Jalbout and E.~I.~Proynov, J.~Math.~Chem.~ \textbf{44}, 260 (2008). 
\bibitem{roy11} A.~K.~Roy, in \emph{Mathematical Chemistry}, W.~I.~Hong (Ed.), Nova Science Publishers, 
Hauppauge, NY, USA, pp.~555-599 (2011). 
\bibitem{roy13} A.~K.~Roy, Int.~J.~Quant.~Chem.~\textbf{113}, 1503 (2013).
\bibitem{varshni90} Y.~P.~Varshni, Phys.~Rev.~A \textbf{41}, 4682 (1990). 
\bibitem{rogers70} F.~J.~Rogers, H.~C.~Graboske Jr.~ and J.~Harwood, Phys.~Rev.~A \textbf{1}, 1577 (1970). 
\end{thebibliography}
\end{document}